# Superconductivity, charge ordering and structural properties of $\alpha, \beta$-$Na_xCoO_2 \cdot y(H_2O, D_2O)$

Y. G. Shi, H. X. Yang, H. Huang, X. Liu, and J.Q. Li*

Beijing National Laboratory for Condensed Matter Physics, Institute of Physics, Chinese Academy of Sciences, Beijing 100080, China

Two series of α- and β-$Na_xCoO_2$ materials have been prepared by means of Na deintercalation from the parent α-$NaCoO_2$ and β-$Na_{0.6}CoO_2$ phases, respectively. The α-$Na_xCoO_2$ materials undergo a clear phase transition from the hexagonal to the β-phase like monoclinic structure along with Na deintercalation. Measurements of resistivity and magnetization demonstrated the presence notable charge ordering transitions in both α- and β-$Na_xCoO_2$ with $0.4 < x < 0.5$ below 100K. Bulk Superconductivity has been observed in the hydrated α-$Na_xCoO_2 \cdot 1.3H_2O$ at ~4.5K and in β-$Na_xCoO_2 \cdot 1.3H_2O$ at ~4.3K. Intercalation of $D_2O$ in α, β-$Na_{0.33}CoO_2$ also yields superconductivity at slightly lower temperatures. It is worthy to note that, despite of the structural difference, the α-, β- and γ-$Na_xCoO_2$ .$yH_2O$ materials show up notable commonalities in their essential physical properties, e.g. superconductivity and charge ordering transitions.

Author to whom correspondence should be addressed: ljq@ssc.iphy.ac.cn.

Layered $\gamma$-$Na_xCoO_2$ system have been investigated systematically in the past years due to its particular properties of large thermoelectric power coexisting with low electric resistivity [1-4]. Recently, superconductivity in the water intercalated $Na_{0.3}CoO_2 \cdot 1.3H_2O$ ($\gamma$-phase) and complex charge-ordering (CO) transitions in $\gamma$-$Na_{0.5}CoO_2$ has been extensively investigated and discussed in connection with the strong electron correlation in present system [5-12]. Actually, there are three distinctive structural series of $Na_xCoO_2$ materials corresponding with the $CoO_6$ octahedra stacking differently along c-axis direction, so called $\alpha$-, $\beta$- and $\gamma$-phase respectively. In order to know the structural and physical properties, especially the superconductivity in $\alpha$, $\beta$-$Na_xCoO_2$ phases, we have performed an extensive study on samples with a variety of Na contents. Actually, a few works, concerning superconductivity in the hydrated $\alpha$- and $\beta$-$Na_xCoO_2 \cdot yH_2O$ samples, have reported that $\alpha$-phase is a superconductor with $T_c \sim 4.6$ K [13] and, on the other hand, the water-intercalated $\beta$-$Na_xCoO_2 \cdot yH_2O$ compound is likely to be not a superconductor [14]. In this paper, we will perform a systematical study on the $\alpha$- and $\beta$-$Na_xCoO_2$ materials, certain notable properties are specially analyzed in comparison with the results obtained from $\gamma$-phase, i.e. structural transformation induced by Na-deintercalation, CO phenomenon at $x \cong 0.5$, and superconductivity in the hydrated $\alpha$- and $\beta$-$Na_xCoO_2$ materials.

Polycrystalline samples of $\alpha$-$NaCoO_2$ were prepared by a conventional solid-state reaction [15]. Powdered cobalt (Co) metal (99.5%) and anhydrous NaOH pellets (Aldrich) in 10 molar excess (Na : Co = 1.1 : 1) were ground together under inert atmosphere and placed in an alumina boat under flowing $O_2$ for approximately 6 days at 500°C with one intermittent grinding. The parent material with nominal composition of $\beta$-$Na_{0.6}CoO_2$ was prepared by a similar process as described above for the $\alpha$-$NaCoO_2$: Co powder and NaOH pellets were mixed in a molar ratio of Na : Co = 0.7 : 1 and ground together under flowing $O_2$ for approximately 5 days at 550°C with one intermediate grinding. Materials with lower Na contents were prepared by the sodium deintercalation [10]. Superconducting samples were synthesized by means of the Na-deintercalation to x ~ 0.3 and then the water intercalation as reported for the preparation of superconducting $\gamma$-$Na_xCoO_2\cdot1.3H_2O$ materials [5]. X-ray diffraction (XRD) measurements were carried out with a diffractometer in the Bragg-Brentano geometry using Cu Ka radiation. The compositions of all materials have been measured by inductively coupled plasma (ICP) analysis technique. Low-temperature magnetization measurements as a function of temperature were performed using a commercial Quantum Design SQUID. Transmission-electron microscopy (TEM) observations were performed on a H-9000NA TEM operating at the voltage of 300kV.

We first focus our attention on the structural evolution of the two series of samples prepared respectively from $\alpha$-NaCoO2 and $\beta$-$Na_{0.6}CoO_2$ by the Na-deintercalation then the water intercalation for samples at around x = 0.3. Fig. 1(a) shows XRD patterns of the $\alpha$-$Na_xCoO_2$ samples for several typical Na contents of x = 1.0, 0.48, 0.43 and 0.33. The diffraction pattern from the hydrated superconducting phase $\alpha$-$Na_{0.31}CoO_2.1.3H_2O$ is also shown to illustrate an evident upsurge of c-axis parameter from $H_2O$ intercalation. The parent $\alpha$-$NaCoO_2$ phase has the *O*3 structure containing three $CoO_2$ layers in an unit cell as reported in ref [13, 15, 16, 17], all diffraction peaks in its XRD pattern shown in fig. 1a can be well indexed on an hexagonal cell with the lattice parameters of a= 2.89Å and c= 15.59 Å (space group of $R\bar{3}m$). The strike structural feature noted in our XRD analysis is a remarkable structural transformation arising from Na-deintercalation; Actually, the well-defined hexagonal $\alpha$-$Na_xCoO_2$ phase is found to be stable only for the Na content of 0.9 < x <1, and a monoclinic structural distortion appears evidently in the Na-deintercalated materials as illustrated for the products with lower Na contents (table 1). In order to study the CO phenomenon in the $\alpha$-$Na_xCoO_2$ system, we have made certain attempts to obtain samples with $x \approx 0.5$ by using either $I_2$ or $Br_2$ in the Na deintercalation, ICP analysis indicated that the products have the compositions of $Na_{0.48}CoO_2$ and $Na_{0.43}CoO_2$, they both show up clear CO transitions at low temperatures as discussed in the following context. The XRD patterns of $\alpha$-$Na_{0.48}CoO_2$, different apparently from that of $\alpha$-$NaCoO_2$ in XRD peak positions, corresponds a *P'*3 structure with the space group of *C*2/m and lattice parameters of a= 4.92Å, b= 5.63 Å, c= 17.36 Å and $\beta$= 105.98°. Another product $\alpha$-$Na_{0.43}CoO_2$ has the same average structure with slightly different lattice parameters as demonstrated in the XRD pattern in fig. 1a. The Na deintercalated material $\alpha$-$Na_{0.31}CoO_2$ is found to be highly sensitive to moisture and shows up visible hydrated feature quickly at the ambient atmosphere as illustrated in fig. 1a; the addition peaks as indicated by asterisks arise actually from the hydrated phase $Na_{0.31}CoO_2 \cdot 0.7H_2O$. This material was further washed in water and stored in a humidified atmosphere for about two days, we finally obtained the superconducting phase $\alpha$-$Na_{0.31}CoO_2 \cdot 1.3H_2O$. This phase has a *P'*3-type structure (space group of $R\bar{3}m$) with the lattice parameters of a= 2.82 Å, and c= 29.57 Å [13].

Fig. 1(b) shows the XRD patterns for the $\beta$-$Na_xCoO_2$ series of samples with different Na contents and $H_2O$ intercalations. The parent sample $\beta$-$Na_{0.6}CoO_2$, as reported in the previously literatures [14, 15, 18], has a *P'*3 structure (space group: C2/m). Systematical analysis on the XRD patterns for the $\beta$-$Na_xCoO_2$ materials, and also in comparison with our above investigation on the $\alpha$-$Na_xCoO_2$ materials, suggest that materials prepared from the either $\alpha$-$NaCoO_2$ or $\beta$-$Na_{0.6}CoO_2$ have an identical structural properties for $x < 0.5$. Hence, our experimental results directly suggest that the *O*3-type hexagonal structure for the parent α-$NaCoO_2$ phase is only stable in the high Na content range in agreement with the previous reported data [15]. In order to facilitate our analysis in present paper, we will still separately discuss the two series of samples obtained respectively from the α-$NaCoO_2$ and β-$Na_{0.6}CoO_2$. In fig. 2b, we show the XRD patterns for two typical samples of β-$Na_{0.41}CoO_2$ and β-$Na_{0.29}CoO_2$, it can be clear recognized that the chief reflection peaks in the XRD patterns correspond perfectly with the known monoclinic structure (C2/m). The β-$Na_{0.29}CoO_2$ sample, as illustrated above for α-$Na_{0.31}CoO_2$, is also easily hydrated at the ambient condition and gives rise to clear peaks from the hydrated phase as indicated by asterisks in fig. 1b.

In order to study the CO phenomenon in α-and β-$Na_xCoO_2$ materials, we have carefully prepared a number of samples with $x \sim 0.5$ by the Na-deintercalation, our experimental results demonstrated that all samples of α- and β-$Na_xCoO_2$ with $0.4 < x < 0.5$ undergo a series of low-temperature phase transition as reported for the CO phase γ-$Na_{0.5}CoO_2$ [10]. Fig. 2(a) shows the XRD patterns taken respectively from the α-$Na_{0.48}CoO_2$ phase in comparison with that of the γ-$Na_{0.5}CoO_2$ phase which was extensively discussed in previous literatures [19]. The α-$Na_{0.48}CoO_2$ compound has a monoclinic unit cell (*P*3 structure) and γ-$Na_{0.5}CoO_2$ sample has an orthorhombic lattice (*P*2 structure), these structural distinctions therefore are clearly recognizable on the XRD peaks appearing at the diffraction angles of larger than $30^o$. In order to examine the microstructure feature, in particular the Na ordering in the CO phase of α- and β-$Na_xCoO_2$, we have performed TEM observations on several typical samples with notable CO transitions. Fig. 2 (b) shows the [001] zone-axis diffraction pattern taken from the α-$NaCoO_2$ sample, illustrating the well-defined hexagonal structure of basic plane. Electron diffraction pattern obtained on the β-$Na_{0.6}CoO_2$ crystal is the same as fig.2 (b) in which no clearly superstructure spots are observed. On the other hand, careful TEM observations on either α-$Na_{0.5}CoO_2$ (0.4 < x < 0.5) or β-$Na_xCoO_2$ (0.4 < x < 0.5) samples reveal very similar systematic satellite spots within the a*-b* plane. Fig. 3 (c) shows an electron diffraction pattern taken from the sample of α-$Na_{0.48}CoO_2$, exhibiting the presence of superstructure spots. This superstructure in general appears as an incommensurate structural modulation with the wave vector of $q = \langle 110 \rangle / 4 + \delta$ ($0 \leq \delta < 0.2$). This kind of structural modulation as previously reported in the γ-$Na_{0.5}CoO_2$ material can be well understood by the zigzag type of Na ordering among the $CoO_2$ sheets [19].

We now proceed to discuss the electric transport and magnetic properties of α- and β-$Na_xCoO_2$ materials in connection with CO transitions that were systematically discussed at low temperatures in the γ-$Na_{0.5}CoO_2$ sample [10, 19]. Fig. 3(a) and (b) show respectively the temperature dependences of resistivity (ρ) and magnetic susceptibility for α-$Na_{0.43}CoO_2$ samples. The α-$NaCoO_2$ phase is a semiconductor with a remarkable resistance increase below 50K [10]. Experimental measurements of resistivity and magnetic susceptibility suggest that all samples of α-$Na_xCoO_2$ (0.4 < x < 0.5) show up the similar low-temperature anomalies as discussed γ-$Na_{0.5}CoO_2$. Fig. 3(c) shows the resistivity ρ of β-$Na_{0.6}CoO_2$ and β-$Na_{0.4}CoO_2$. The β-$Na_{0.6}CoO_2$ sample is metallic, its resistivity decreases linearly with lowering temperature. The β-$Na_{0.41}CoO_2$ sample, similar with the other $x \approx 0.5$ materials, shows up a sharp upturn in resistivity at the temperature of around 53 K, a notable saturation just below 25 K following by another increase of the slope. In general, magnetization measurements (see fig. 3b and fig. 3d) on either α- or β-$Na_xCoO_2$ samples with 0.4 < x < 0.5 clearly revealed the presence of three phase transitions at the temperatures of around 25 K, 50 K and 90 K respectively. The 53K-transition was proposed to be associated with a notable magnetic ordering of two distinctive Co sites, and, therefore, an remarkable upturn of resistivity appears in general accompany this transition [10]. This fact also demonstrates that the correlated alternations of CO and magnetic structure indeed exist in present materials at low temperatures. The transition at 90 K is proposed in connection with certain kind of structural changes [19]. It is also noted the CO materials of α- and β-$Na_xCoO_2$, in contrast with the known CO γ-$Na_{0.5}CoO_2$ phase, often have a notable antiferromagnetic (AFM) background as seen in fig.3b and d. We firstly considered this kind of AFM signal arises from an impurity phase, we therefore make a lot of effort to increase the sample quality, however, this AFM background remains visible in the samples in which TEM and XRD structural analysis cannot detect any impurity phases. Another possible original of this AFM background arises from the phase separation commonly appearing in the sample with x ~ 0.5 [19], i.e. certain fraction of materials in the CO materials with different Na content possibly produce complex AFM property.

Superconductivity in the α- and β-$Na_xCoO_2$ materials are another key issue concerned in present study. Structural investigation on the hydrated samples demonstrated the presence of certain typical hydrated phases with nominal compositions of α- and β-$Na_{0.3}CoO_2 \cdot yH_2O$ (y=0.7, and 1.3) similar with what observed in the γ-$Na_{0.3}CoO_2 \cdot yH_2O$. The supercomputing phase, with the nominal compositions of α- and β-$Na_{0.3}CoO_2 \cdot 1.3H_2O$, have a distance of ~9.9Å between two adjacent $CoO_2$ sheets in consistence with the data of the known γ-$Na_{0.33}CoO_2 \cdot 1.3H_2O$ superconductor. Figure 4 (a) shows the zero-field cooling (ZFC) magnetization data measured in a field of 20Oe for α- and β-$Na_{0.3}CoO_2 \cdot 1.3H_2O$ samples. The presence of strong diamagnetic signals for both samples provides direct evidence for bulk superconductivity in these two hydrated phases. The superconducting transitions, changing slightly from one sample to another, occurs at around $Tc_{onset}$~4.2 K for α-$Na_{0.3}CoO_2·1.3H_2O$ and $Tc_{onset}$~4.3 K for β-$Na_{0.3}CoO_2·1.3H_2O$. In previous literatures, the bulk superconductivity was observed in the hydrated α-$Na_{0.3}CoO_2·1.3H_2O$ at about *Tc* ~4.6 K [13]. On the other hand, no bulk superconductivity was detected in the hydrated β-phase [14]. Careful analysis on α- and β-$Na_{0.3}CoO_2·1.3H_2O$ materials suggests that the existence of a small fraction of impurity phases often totally destroy superconductivity in this kind of system. Hence, the acquirements of superconducting phases in either α- or β-$Na_xCoO_2·yH_2O$ materials are found to be more difficult than that in the γ-$Na_{0.3}CoO_2·1.3H_2O$ phase.

Alternations of intercalated layers among the $CoO_2$ sheets are expected to have certain effects on the structure as well as superconductivity as discussed in the γ-$Na_{0.3}CoO_2·1.3H_2O$ and γ-$Na_{0.3}CoO_2·1.3D_2O$ superconducting phases [7]. In α, β-$Na_xCoO_2$ systems, we have prepared a series of the $D_2O$-intercalated samples with nominal composition of α-$Na_{0.3}CoO_2·1.3D_2O$. X-ray diffraction result as shown in the insert of Fig. 4 (b) demonstrates that this phase has a similar hexagonal structure with the $H_2O$-intercalated superconducting phase, the only recognizable change is the decrease of c-axis parameter from 29.56 Å to 29.42 Å. Fig. 4 (b) shows the temperature dependence of the magnetic susceptibility of a α-$Na_{0.3}CoO_2·1.3D_2O$ sample showing a superconducting transition at around 3.7 K, the data for the α-$Na_{0.3}CoO_2·1.3H_2O$ is also shown for comparison. Our systematic analysis on the $D_2O$ intercalated samples suggests that the substitution of $D_2O$ for $H_2O$ is likely to lower the superconducting transition. However, considering the structural complexity in this kind of materials, numerous factors, such as the Na content, the $H_2O$ ($D_2O$) layer structures and small fraction of impurities, might has notable effects on superconductivity. Hence, we cannot conclude the essential role of the $D_2O$ substituting for $H_2O$ on superconductivity ($T_c$) based on present experimental data, a further study performing on several high quality samples are still in progress.

In summary, we have successfully prepared two series of α- and β-$Na_xCoO_2$ materials by means of Na deintercalation from the parent α-$NaCoO_2$ and β-$Na_{0.6}CoO_2$ phases, respectively. The α-$Na_xCoO_2$ materials undergo a clear phase transition from the hexagonal to the β-phase like monoclinic structure along with Na deintercalation, as a result, α- and β-$Na_xCoO_2$ materials is likely to have the identical crystal structure for sample with $x < 0.5$. Measurements of resistivity and magnetization demonstrated the presence notable CO transitions in all α- and β-$Na_xCoO_2$ materials with $0.4 < x < 0.5$, these transitions is very similar with those observed in the γ-$Na_{0.5}CoO_2$ phase. Bulk superconductivity has been found in the hydrated α-$Na_{0.3}CoO_2 \cdot 1.3H_2O$ at ~4.2K and in β-$Na_xCoO_2 \cdot yH_2O$ at ~4.3K. Intercalation of $D_2O$ in α, β-$Na_{0.33}CoO_2$ also yields superconductivity at 3.7K. Despite of the structural difference, the α-, β- and γ-$Na_xCoO_2$ materials show up notable commonalities in superconductivity and CO transitions.

**Acknowledgments**

We would like to thank Mr. W. W. Huang, Miss. H. Chen and Miss. S. L. Jia for their assistance in preparing samples and measuring some physical properties. The work reported here is supported by National Natural Foundation and by the 'Outstanding Youth Fund' (JQL) with Grant No 10225415 of China.

**Figure and Table Captions**

Table 1. Structural properties and ICP results for the typical α- and β-$Na_xCoO_2$ .$yH_2O$ materials.

Figure 1. XRD patterns for (a) α-$Na_xCoO_2$, (b) β-$Na_xCoO_2$, illustrating the structural alternations along with Na content and $H_2O$ intercalation.

Figure 2. (a) Comparison of the XRD patterns between the charge-ordered α-$Na_{0.48}CoO_2$ and γ-$Na_{0.5}CoO_2$, the structural distinction can be clearly recognized. The [001] zone axis electron diffraction pattern taken from (b) α-$NaCoO_2$ and (c) α-$Na_{0.48}CoO_2$, the presence of superstructure spots in the charge ordered phase is evident.

Figure 3. Temperature dependence of magnetic susceptibility (χ) and resistivity (ρ) for (a, b) the α-$Na_{0.43}CoO_2$ and (c, d) the β-$Na_{0.41}CoO_2$ materials, clearly demonstrating the low-temperature charge-ordering transitions.

Figure 4. (a) The magnetic susceptibilities for the α- and β-$Na_{0.3}CoO_2 \cdot 1.3H_2O$ superconducting samples. (b) The magnetic susceptibilities for the α-$Na_{0.3}CoO_2 \cdot 1.3D_2O$, shwing the superconducting transition at 3.7 K. The data from α-$Na_{0.3}CoO_2 \cdot 1.3H_2O$ are cited for comparison.

Table 1

| Samples | ICP data (Na : Co) | Space group | Lattice Parameters (Å) |
|---|---|---|---|
| α-parent sample | 0.99 | $R\bar{3}m$ | Hexagonal<br>a= 2.89, c= 15.59 |
| α-CO samples | 0.48 | C2/m | Monoclinic<br>a = 4.92, b= 5.63, c= 17.36, β= 105.98 |
| | 0.43 | C2/m | Monoclinic<br>a= 4.92, b= 5.64, c= 17.39, β= 105.76 |
| α-superconducting sample | 0.31 | $R\bar{3}m$ | Hexagonal<br>a= 2.82, c= 29.57 |
| β-parent sample | 0.63 | C2/m | Monoclinic<br>a= 4.89, b= 5.65, c= 17.14, β= 106.18 |
| β-CO sample | 0.41 | C2/m | Monoclinic<br>a= 4.92, b= 5.65, c= 17.39, β= 106.10 |
| β- superconducting sample | 0.29 | $R\bar{3}m$ | Hexagonal<br>a= 2.82, c= 29.56 |

Figure 1

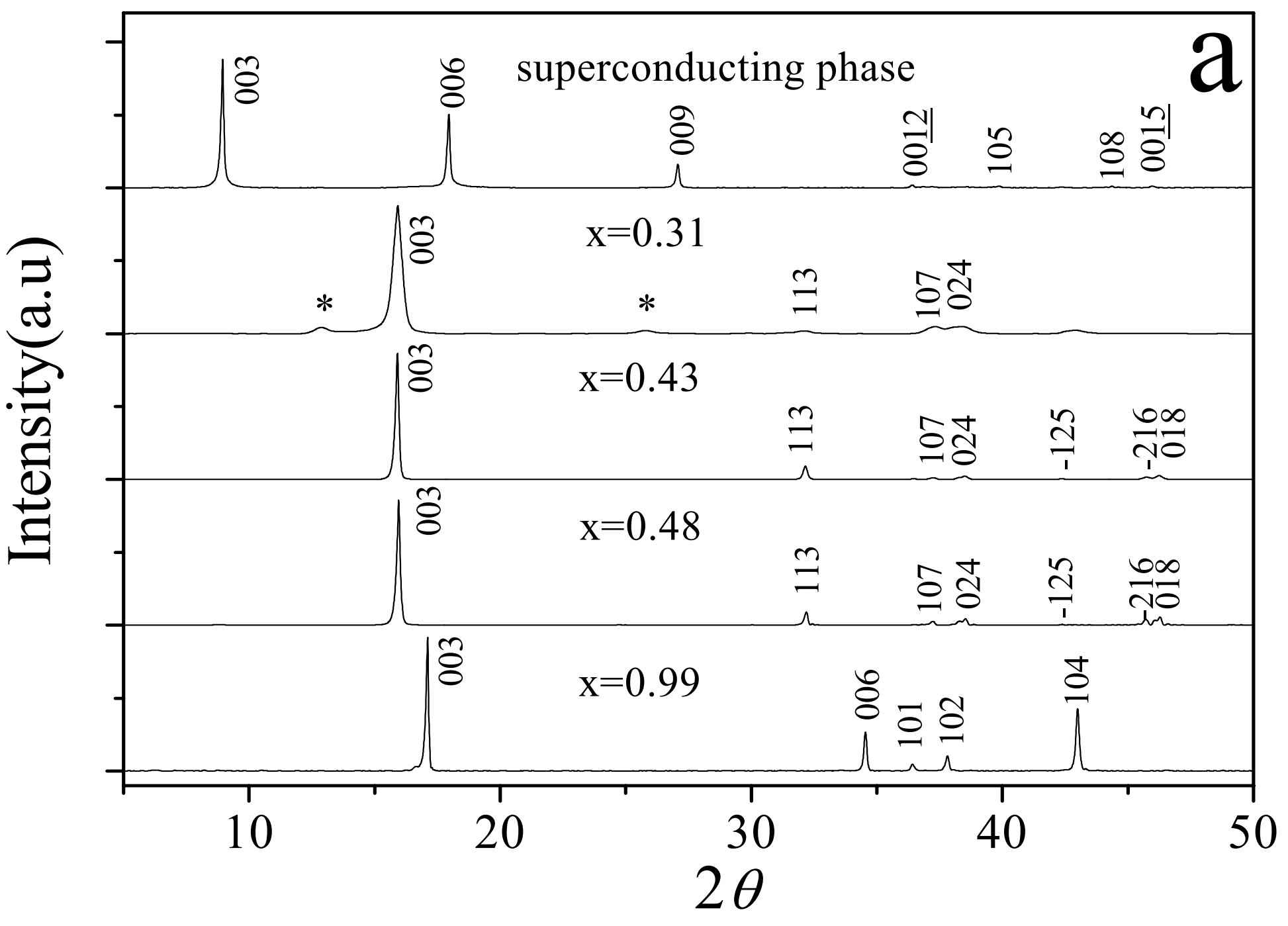

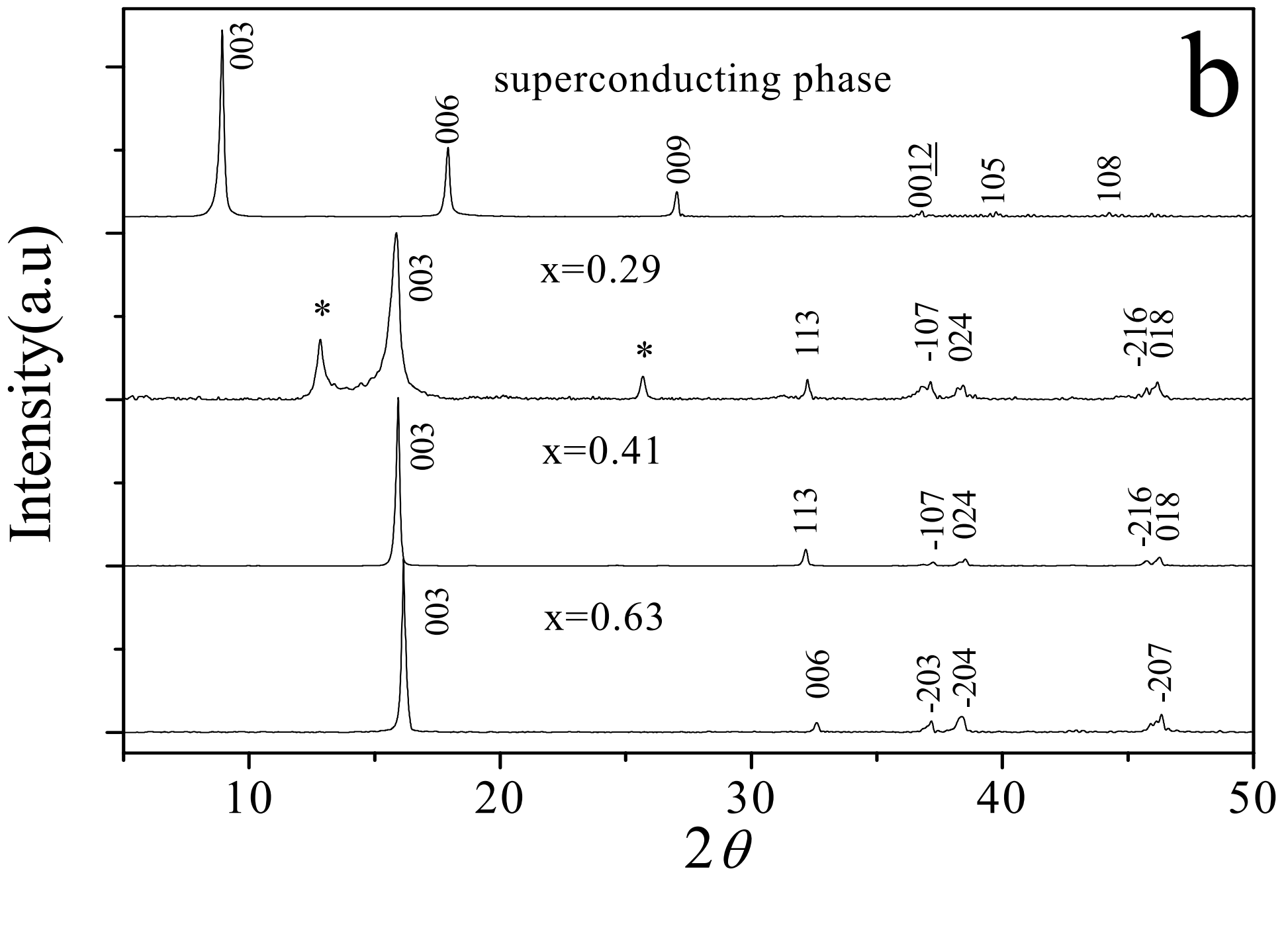

Figure 2

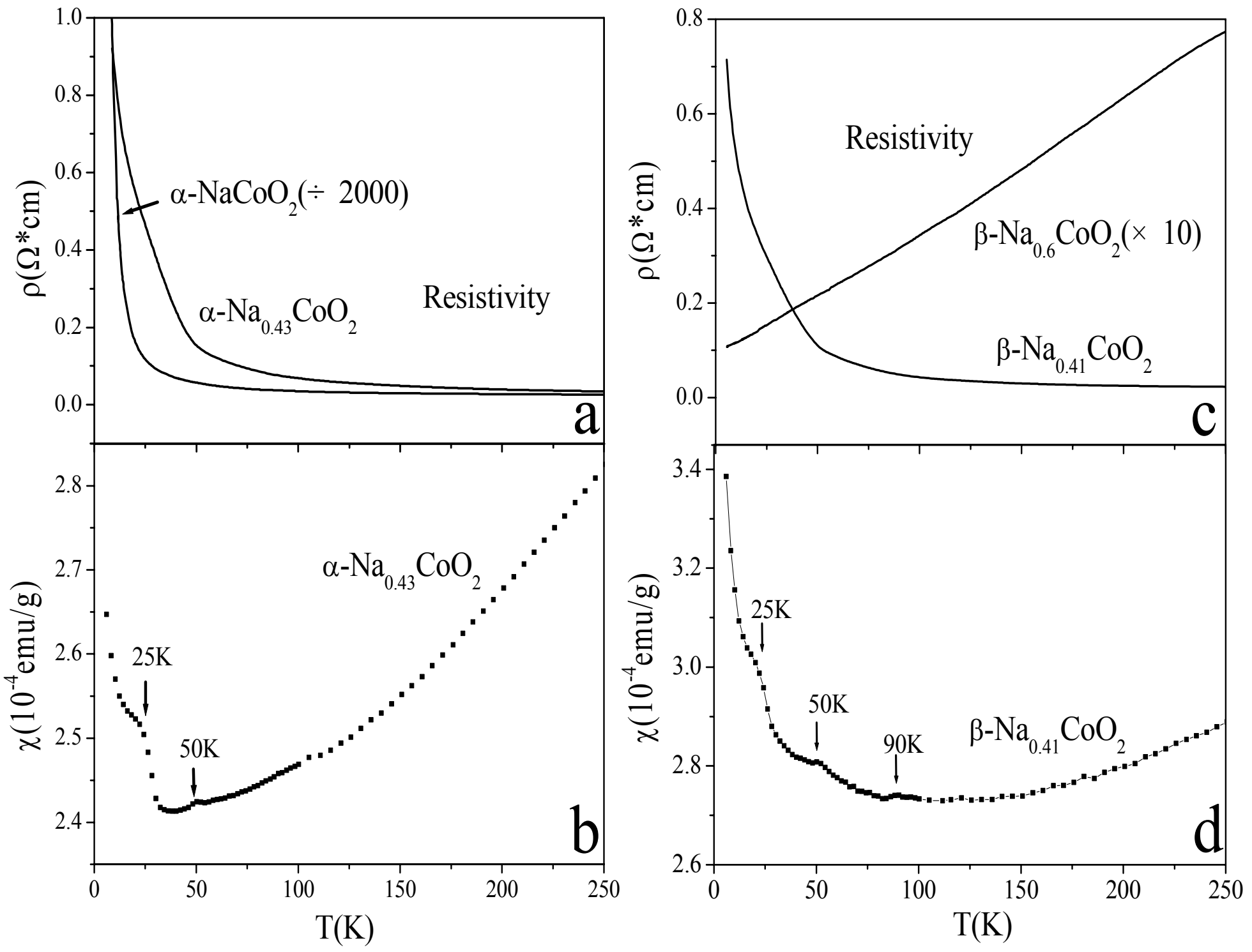

Figure 3

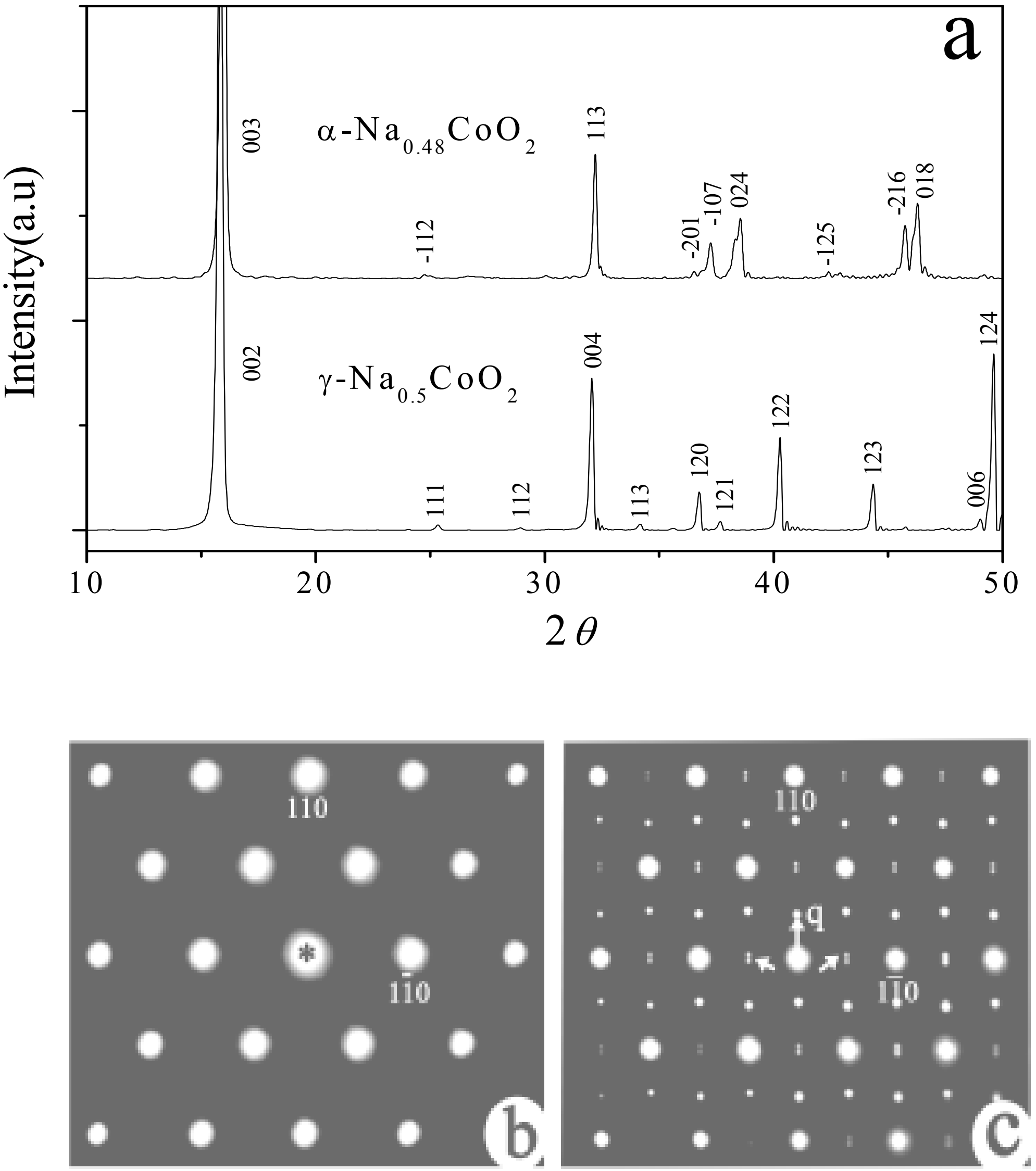

Figure 4

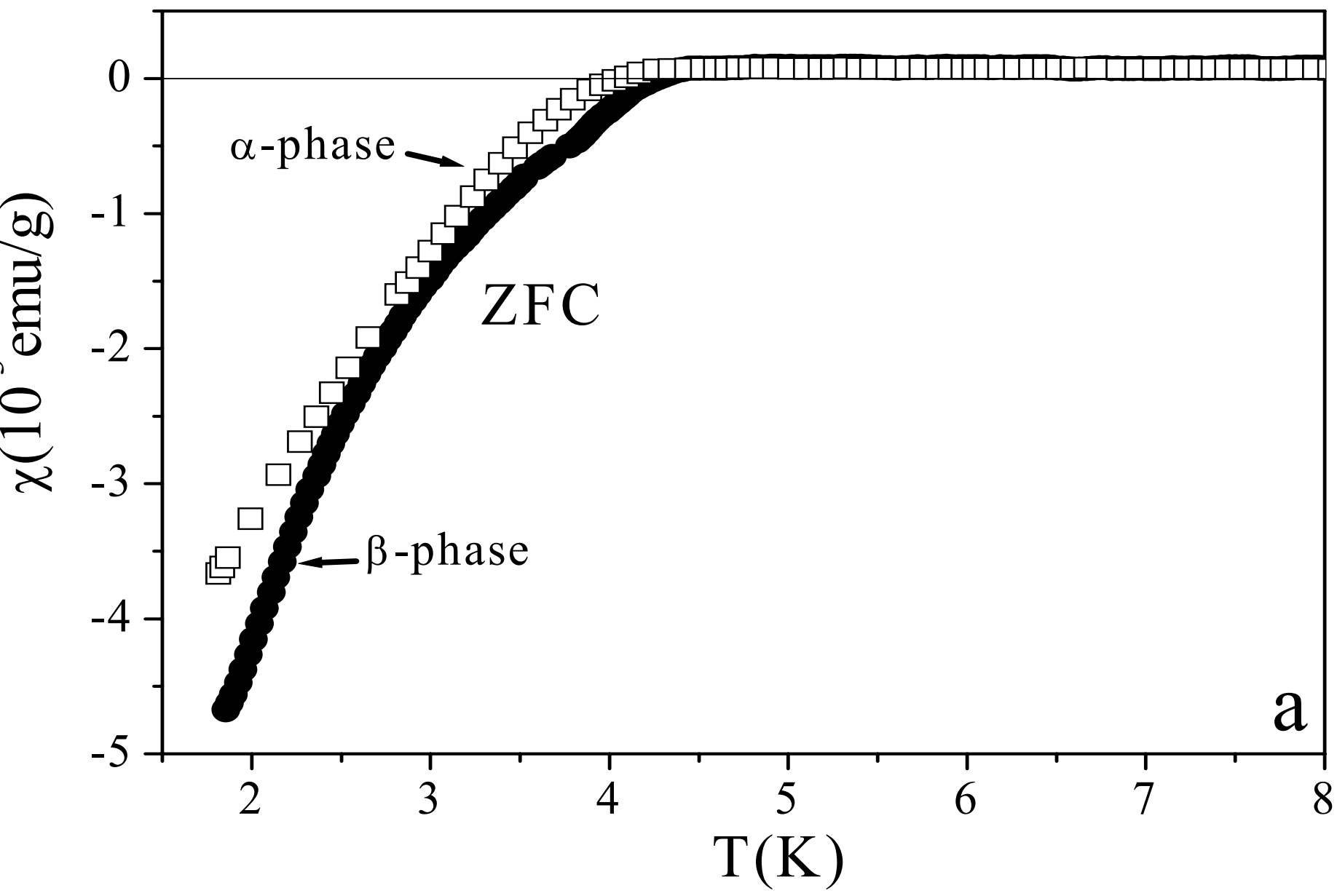

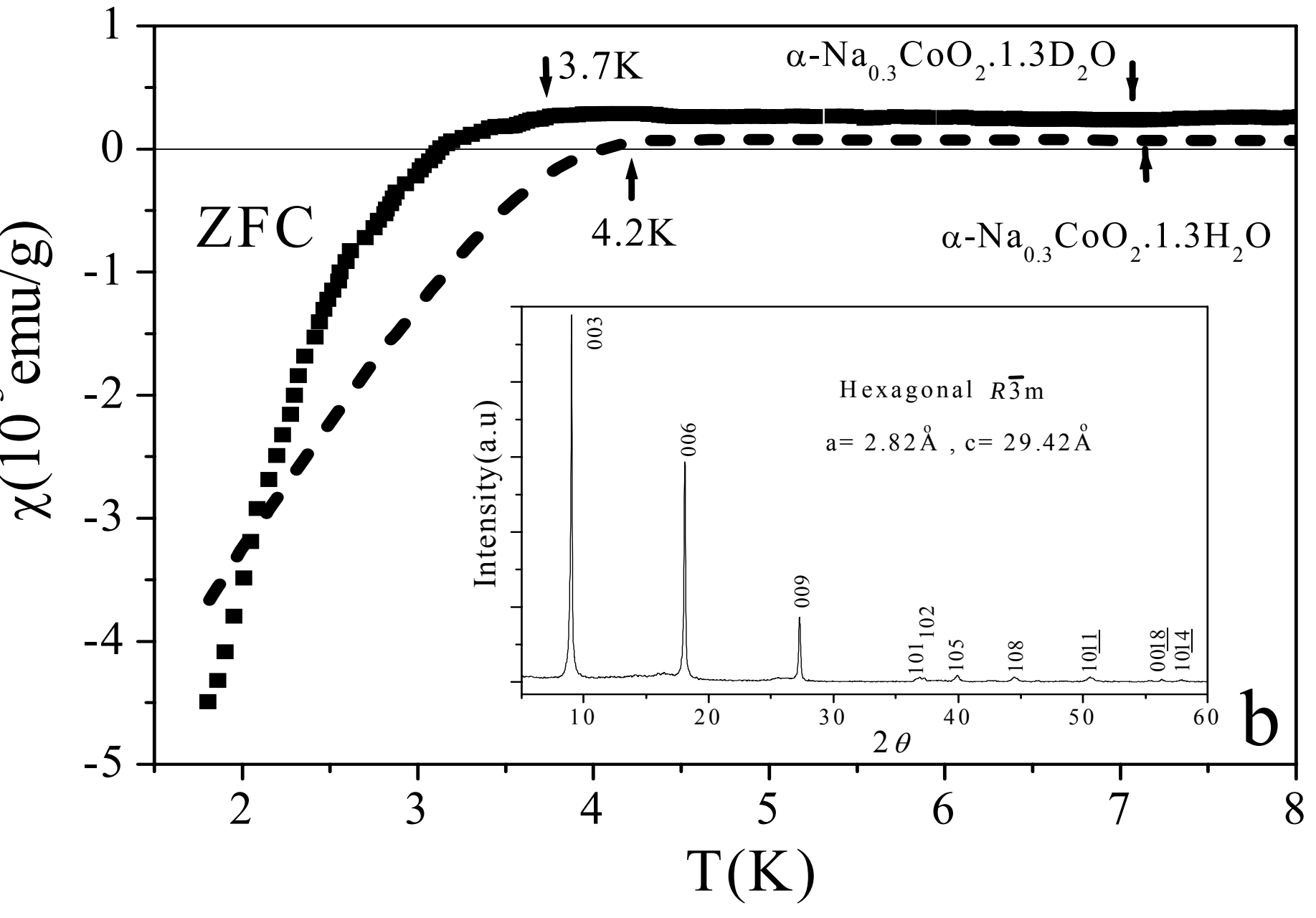